\documentclass{article}

\usepackage{arxiv}

\usepackage[utf8]{inputenc} 
\usepackage[T1]{fontenc}    
\usepackage{hyperref}       
\usepackage{url}            
\usepackage{booktabs}       
\usepackage{amsfonts}       
\usepackage{nicefrac}       
\usepackage{microtype}      
\usepackage{lipsum}		
\usepackage{graphicx}
\usepackage{siunitx}
\usepackage{subcaption}
\usepackage{amsmath}
\usepackage{amssymb}

\title{Agent-based Simulation of Pedestrian Dynamics for Exposure Time Estimation in Epidemic Risk Assessment}

\author{Thomas Harweg$^{\hspace{1mm}\href{https://orcid.org/0000-0001-8477-481X}{\includegraphics[scale=0.06]{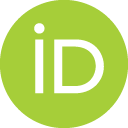}}}$  \\
        Department of Computer Science \\
        TU Dortmund University\\
	\texttt{thomas.harweg@tu-dortmund.de} \\
	\And
Daniel Bachmann$^{\hspace{1mm}\href{https://orcid.org/0000-0002-8767-0669}{\includegraphics[scale=0.06]{orcid.png}}}$ \\
        Department of Computer Science \\
        TU Dortmund University\\
	\texttt{daniel.bachmann@tu-dortmund.de} \\
	\And
Frank Weichert$^{\hspace{1mm}\href{https://orcid.org/0000-0002-2530-8197}{\includegraphics[scale=0.06]{orcid.png}}}$ \\
        Department of Computer Science \\
        TU Dortmund University\\
	\texttt{frank.weichert@tu-dortmund.de} \\
}

\renewcommand{\headeright}{}
\renewcommand{\undertitle}{}

\begin{document}
\maketitle

\begin{abstract}
With the Corona Virus Disease 2019 (COVID-19) pandemic spreading across the world, protective measures for containing the virus
are essential, especially as long as no vaccine or effective treatment is available.
One important measure is the so-called \emph{physical distancing} or \emph{social distancing}.
In this paper, we propose an agent-based numerical simulation of pedestrian dynamics in order to
assess behaviour of pedestrians in public places in the context of contact-transmission
of infectious diseases like COVID-19, and to gather insights about exposure times and the overall effectiveness of distancing measures.
To abide the minimum distance of \SI{1.5}{\metre} stipulated by the German government at an infection rate of 2\%, our simulation results suggest that
a density of one person per \SI{16}{\square\metre} or below is sufficient.
The results of this study give insight about how physical distancing as a protective measure can be carried out more
efficiently to help reduce the spread of COVID-19.
\end{abstract}


\section{Introduction}
\label{sec:introduction}
Starting at the end of 2019, the Corona Virus Disease 2019 (COVID-19) was first described in Wuhan, China~\cite{Zhou:2020}, and has rapidly spread world-wide
over the past months causing an unprecedented pandemia with more than 431, 541 deaths so far (first wave of disease) \cite{WHO2020}. The illness is caused by
the Severe Acute Respiratory Syndrome Corona Virus 2 (SARS-CoV-2). Despite drastic restrictions in every-day life, numbers of new infections are still on the
rise~\cite{Dong:2020}. Hence, special attention should be given to the protection of vulnerable patients with high risk of a severe course of the disease.
Diagnostic gold standard to identify SARS-CoV-2 infection is reverse transcription polymerase chain reaction (RT-PCR)
of viral ribonucleic acid (RNA) collected by a combined nasopharyngeal swab (NPS) and oropharyngeal
swab (OPS)~\cite{Zou:2020}. As long as no vaccine or at least no therapy is available, only exit restrictions and social distancing can slow down the spread
of COVID-19. Social distancing, also called ``physical distancing'' means limiting face-to-face contact with others. A model developed to support pandemic
influenza planning~\cite{Ferguson:2006, Halloran:2008} was adapted using the data of the COVID-19 outbreak in Wuhan to explore scenarios for the United States and
Great Britain~\cite{Ferguson:2020} resulting in the advice of social distancing of the whole population and household quarantine of infected individuals as well as
school and university closures. By simulating the COVID-19 outbreak in Wuhan using a deterministic stage-structured SEIR (susceptible, exposed, infectious, recovered)
model over a 1 year period,
\emph{Prem et al.}~\cite{Prem:2020} came to the conclusion that a reduction in social mixing can be effective in reducing magnitude and delaying the peak of
outbreak. These British epidemiologists from the London School of Hygiene \& Tropical Medicine suspect that a second wave of COVID-19 disease could only be
prevented if exit and contact restrictions were maintained over the long term or at least resumed intermittently. It is therefore increasingly important to
know what distances must be maintained to avoid infection. As far as these distances are concerned the study of \emph{Bischoff et al.}~\cite{Bischoff2013}
revealed that healthcare professionals within \SI{1.829}{\metre} of patients with influenza could be exposed to infectious doses of influenza virus, primarily
in small-particle aerosols. This led to the advice of keeping a minimum distance of \SIrange{1}{2}{\metre} from the
\emph{Robert Koch Institut (RKI)}~\cite{RKI2020a},
which is evaluating available information
of the corona virus and estimating the risk for the population in Germany. The RKI and the German Federal Ministry of Health has in response published a
handout~\cite{RKI2020b}
stating that in public, a minimum distance of \SI{1.5}{\metre} must be maintained wherever possible. Besides, it is relevant whether people are next to each
other, staggered behind each other, or directly behind each other. Especially in indoor areas, e.g. shopping, corresponding constellations occur in combination.
Simulations can help to make the necessary distance measures easier to understand. Different distance scenarios can be simulated and recommendations for the
distance can be suggested.\\

In their call to action \emph{Squazzoni et al.}~\cite{Squazzoni:2020} give an overview of computational models for global pandemic outbreak simulation and their
 limitations and \emph{Chang et al.}~\cite{Chang:2020} propose a microscopic model for the simulation of the COVID-19 outbreak in Australia. The model consists
  of over 24 million individuals with different characteristics and social context and was calibrated with the 2016 census data of Australia.\\

In the context of the simulation of pedestrian dynamics there exists a variety of comprehensive surveys \cite{Shiwakoti:2008, Schadschneider:2009, Caramuta:2017}.
Models were differentiated mainly into macroscopic and microscopic. In macroscopic models the crowd is assumed as the smallest entity. These models allow the
representation of high density crowds, but cannot model the interaction between pairs or groups of pedestrians. Microscopic models assume one pedestrian as the
smallest entity. They are roughly divided into physical force models, cellular automata or queueing models. This work focusses on the so called social force
model by \emph{Helbing et al.}~\cite{Helbing:1991, Helbing:2000, Helbing:2001, Helbing:2002a} which can be categorised as a physical force model.\\

Main contributions of this work are
\begin{itemize}
	\item adaptable social force-based model for pedestrian dynamics in realistic environments
	\item exposure time measurement for assessment of the spreading of diseases
	\item discussion of the effects of distancing measures on exposure times.
\end{itemize}

This paper is organised as follows. Firstly, Section~\ref{sec:related_work} gives an overview of existing models for simulations of the spread of diseases.
In Section~\ref{sec:materials_and_methods} we set forth the agent-based model and the accompanying simulation we used for our experiments.
Section~\ref{sec:experiments} describes the experiments concerning the simulation of pedestrian dynamics in realistic environments,
the effectiveness of distancing measures, and measuring of exposure time in the context of infectious diseases in general and COVID-19 in particular.
Following, in Section~\ref{sec:results} we present the results from the experiment, and in Section~\ref{sec:discussion} we give a short conclusion/summary and outlook.

\section{Related Work}
\label{sec:related_work}
As far as the simulation of the spreading of diseases is concerned most approaches are based on macroscopic models. The so called compartment models~\cite{Brauer:2008}
divide the population of interest into compartments with different characteristics. The simplest model is the SIR model. It consists of three compartments
\emph{susceptible}, \emph{infectious} and \emph{recovered}. The population is split into these three compartments. Entities in the susceptible group model the entities
most likely to be infected. The entities in the infectious group are the ones already infected and the entities in the recovered group have recovered from the infection.
An entity transfer from for example susceptible to infectious state and from infectious to recovered state could be modelled. Thus, re-infection with the modelled
disease would not be depicted by this model. A plethora of different variations of this scheme exist with varying numbers of compartments~\cite{Hethcote:2000}.
The independent variable in the compartment models is the time $t$. The transfer ratios of the population from one compartment to another are expressed as derivatives
with respect to $t$, thus resulting in differential equations for the compartments of the model. One of the shortcomings of these models is, that each individual in the
population is modelled with the same set of features. This is overcome by the introduction of metapopulations~\cite{Brockmann:2006, Balcan:2009} building sub compartments
of, e.\,g., entities with natural immunity or asymptomatic individuals. Still each of the metapopulations share a homogeneous set of model parameters.

With rapid growth of available processing power individuals based models (IBM) or agent based models (ABM) are used to model infectious disease outbreaks.
\emph{Willem et al.}~\cite{Willem:2017} and \emph{Nepomuceno et al.}~\cite{Nepomuceno:2019} give comprehensive overviews of IBM/ABM usage in the field of epidemiology.
Based on the work by \emph{Brockmann}~\cite{Brockmann:2010}, \emph{Fr\'ias-Mart\'inez et al.}~\cite{FriasMartinez:2011} propose an agent-based model of epidemic spread
based on social network information from data of base transceiver stations (BTC) captured during the 2009 H1N1 outbreak in Mexico. By simulating an outbreak of
measles that occurred in Schull, Ireland in 2012 based on open data, \emph{Hunter et al.}~\cite{Hunter:2018} have shown recently, that agent based modelling in combination
with open data leads to regionally transferable models.
\emph{Bobashev et al.}~\cite{Bobashev:2007} discuss the combination of compartment models and microscopic agent-based models into an so called hybrid multi-scale model.

In this paper, we propose a microscopic model for simulating pedestrian dynamics in the context of infectious disease spread, including monitoring of contacts and exposure
times in realistic scenarios. The simulation devised in this work runs in real-time, giving instantaneous feedback about the considered scenario and thus allowing for
visual assessment, in addition to resulting statistics.

\section{Materials and Methods}
\label{sec:materials_and_methods}
In the following, we describe in detail the underlying physical model of pedestrian dynamics and the simulation we implemented.
Our simulation is adapted to realistic scenarios in the context of the assessment of infectious disease spread, with focus on exposure time measurement.
Specifically, we apply it to a supermarket scenario and the corresponding measures taken with regard to the COVID-19 pandemic.

The simulation we present is an agent-based approach based on the model of \emph{Helbing et al.}~\cite{Helbing:1991, Helbing:2000, Helbing:2001, Helbing:2002a}.
The simulation is carried out on a static scenery with a defined number $n \in \mathbb{N}$ of \textit{agents} or \textit{particles} $p_i$.
Each agent has an individual starting point $\mathbf{s}_i$ and destination point $\mathbf{d}_i(t)$ (both $\in \mathbb{R}^2$), the former being static and the latter
varying with time $t \in \mathbb{R}_{\geq 0}$.

Motion of an individual $p_i$ is governed by Equation~\ref{equ:motion}, which is composed of the term for self-propelling $\mathbf{a}^{\textrm{self}}$ and external forces
$\mathbf{f}$ acting upon $p_i$
\begin{equation}
\label{equ:motion}
\frac{d\mathbf{x}^{2}_{i}(t)}{dt^{2}} = \mathbf{a}^{\textrm{self}}_{i} + \frac{1}{m_i} \cdot \left(\sum_{j} (\mathbf{f}_{ij}^{\textrm{soc}} + \mathbf{f}_{ij}^{ph})  + \mathbf{f}^{\textrm{wall}}_{i})\right)\textrm{.}	 
\end{equation}
Here, $\mathbf{x}_i\in \mathbb{R}^{2}$ denotes the position of $p_i$, and $m_{i}$ (in \SI{}{\kilogram}) its mass.

Self-acceleration of agents, as defined in Equation~\ref{equ:self_accel}, is the adjustment of the actual velocity $\mathbf{v}_i(t)$ to the desired velocity
\begin{equation}
\label{equ:self_accel}
\mathbf{a}^{\textrm{self}}_{i}(t) = \frac{v^{0}_{i}\mathbf{e}^{0}_{i}(t) - \mathbf{v}_{i}(t)}{\tau}\textrm{.}
\end{equation}
The parameter $\tau \in \mathbb{R}_{>0}$ (in \SI{}{\second}) determines the amount of delay time for an agent to adapt. The desired velocity of an agent $p_i$
is the product of its speed $v^{0}_{i}$ and the desired direction of movement $\mathbf{e}^{0}_{i}(t)$. The latter depends on the destination $\mathbf{d}_i(t)$ and is determined
by the pre-computed navigation method described below.

The model by Helbing mainly consists of different types of forces acting upon an agent, social and physical forces. The social forces $\mathbf{f}^{\textrm{soc}}$ model a pedestrian's endeavour to
avoid contact with other pedestrians, while the physical forces $\mathbf{f}^{\textrm{ph}}$ model effects arising when pedestrians are so close that there is actual contact between them.
While in the original model of Helbing et al.\ the social force is only comprised of a normal component (called $\mathbf{f}^{\textrm{norm}}$ here), in this paper,
we add an additional tangential term $\mathbf{f}^{\textrm{tang}}$.

\begin{figure}[tb]
	\centering
	\begin{subfigure}{0.3\linewidth}
		\centering
		\includegraphics[height=0.1\textheight]{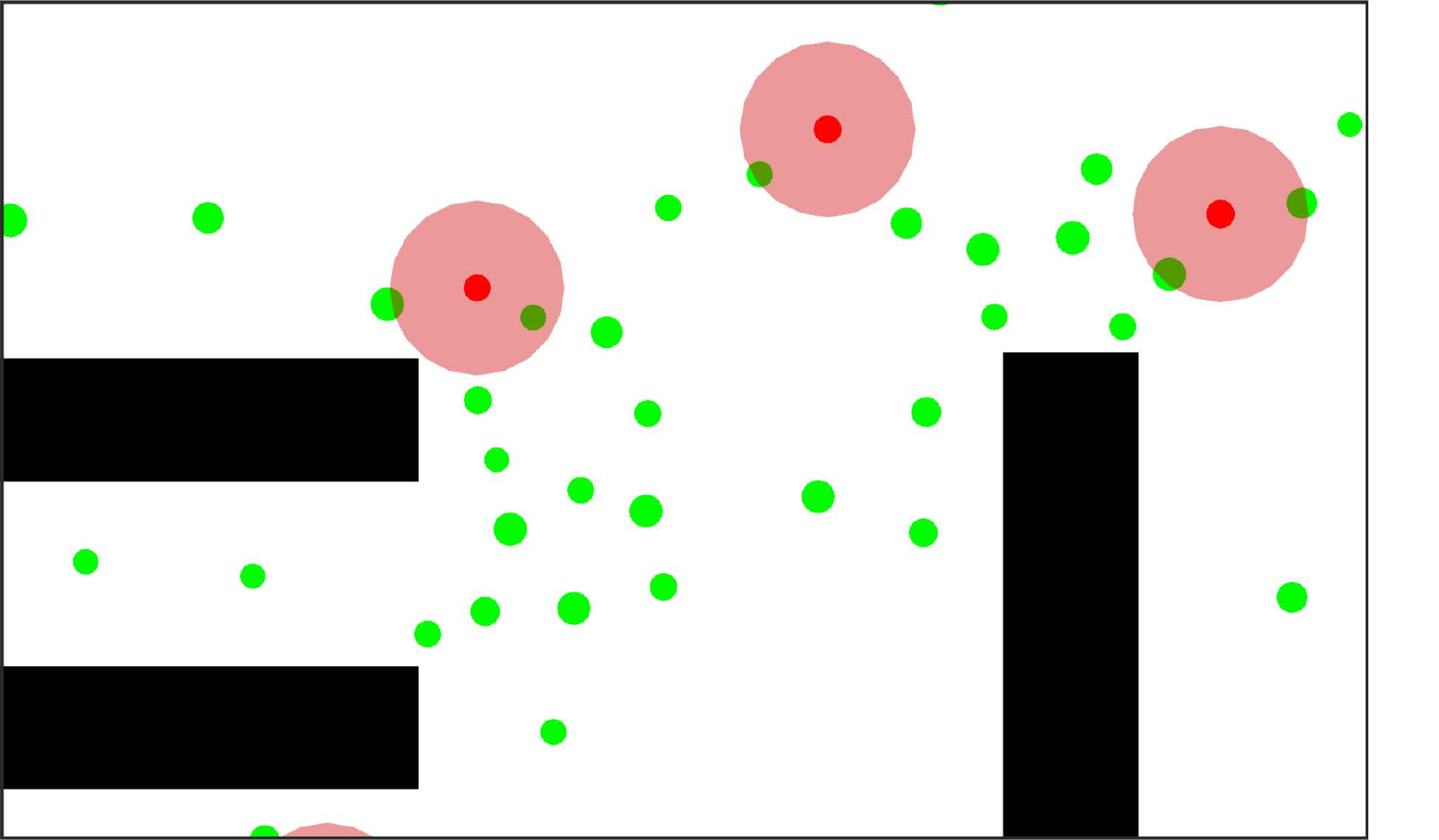}
		\caption{Interaction of agents (infectious and non-infectious) during simulation. Critical radius is shown in red around infectious agents.}
		\label{fig:agent_interact_sim}
	\end{subfigure}
	\hspace{0.2cm}
	\begin{subfigure}{0.3\linewidth}
		\centering
		\includegraphics[height=0.1\textheight]{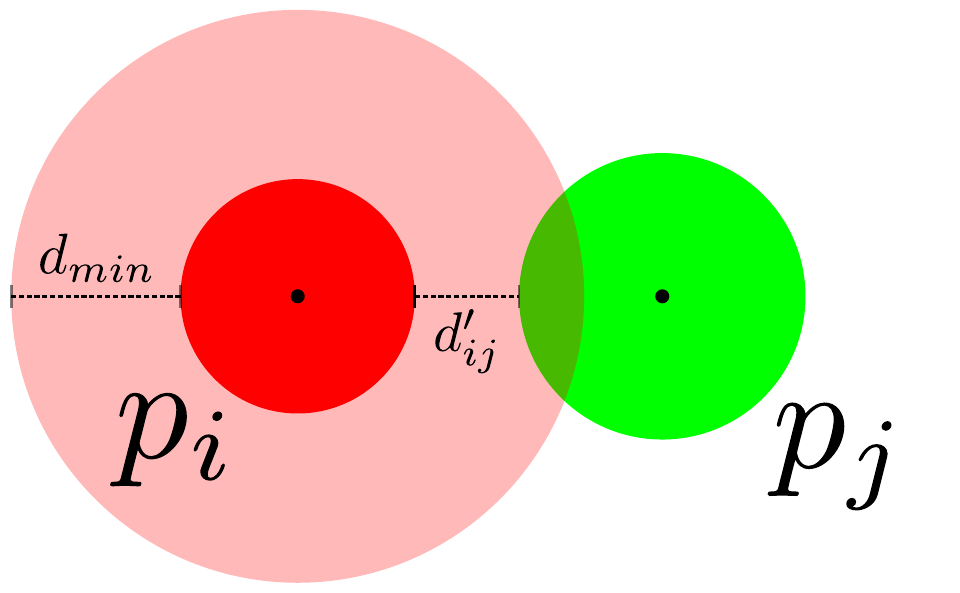}
		\caption{Visualisation of two agents $p_{i}$ and $p_{j}$ coming below the critical distance $d_{min}$}
		\label{fig:agent_distance}
	\end{subfigure}
	\hspace{0.2cm}
	\begin{subfigure}{0.3\linewidth}
		\centering
		\includegraphics[height=0.1\textheight]{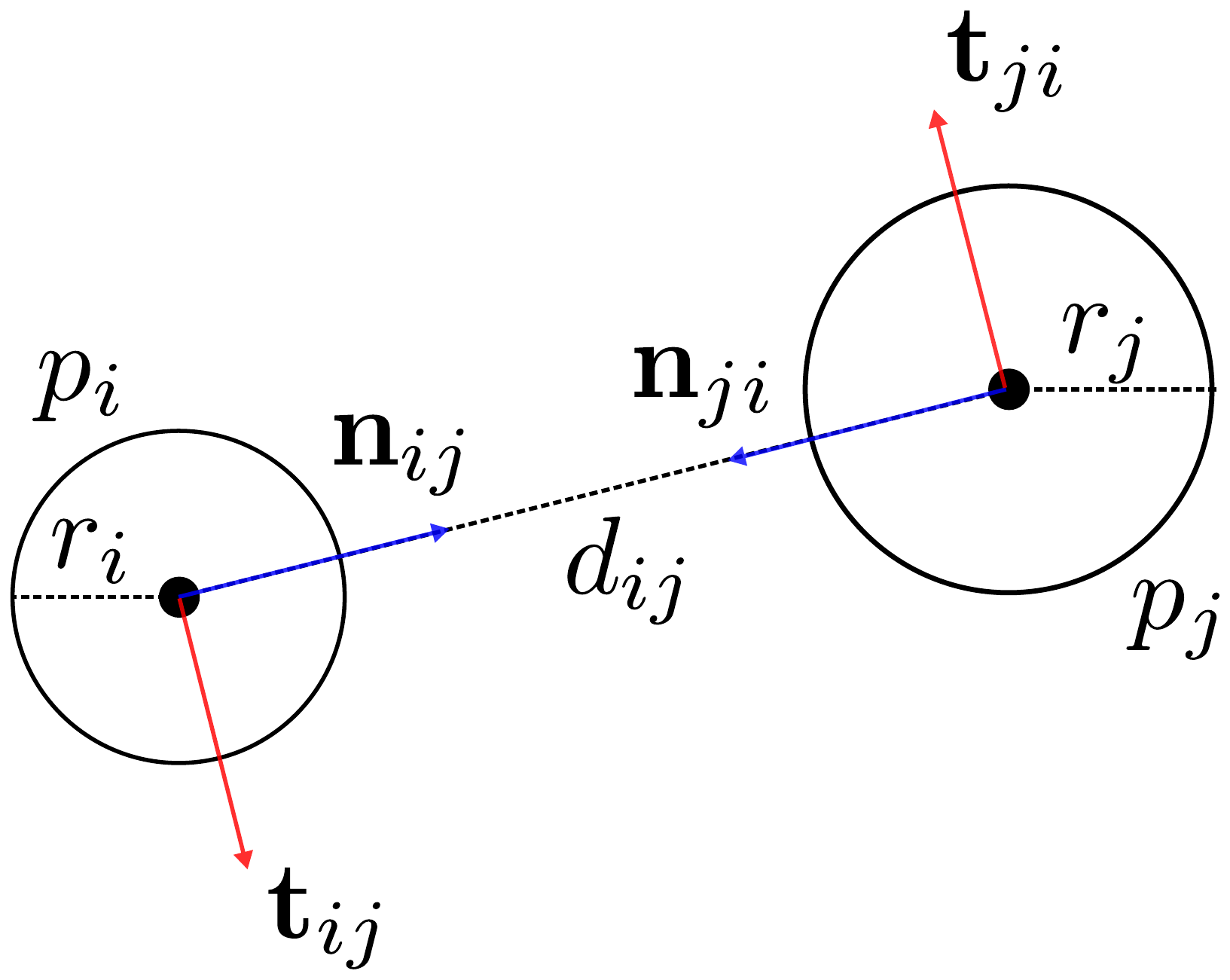}
		\caption{Underlying model of agent interaction between two agents $p_{i}$ and $p_{j}$}
		\label{fig:agent_interact_pi_pj}
	\end{subfigure}
	\caption{Agent interaction in the simulation. Figure (a) shows a visualisation of the running simulation, and (b) shows the measuring of distances.
		Figure (c) shows associated quantities of the underlying physical model}
	\label{fig:agent_interact}
\end{figure}

The social force thus takes the following form:
\begin{equation}
\label{equ:forces_soc}
\mathbf{f}^{\textrm{soc}} = \mathbf{f}^{\textrm{norm}} + \mathbf{f}^{\textrm{tang}}.
\end{equation}

The accompanying force in normal direction is defined as usual~\cite{Helbing:1991, Helbing:2002a}:
\begin{equation}
\label{equ:force_norm}
\mathbf{f}^{\textrm{norm}}_{ij}(t) = A \cdot \exp \left( \frac{r_{ij}-d_{ij}(t)}{B} \right) \cdot \mathbf{n}_{ij}(t)\cdot \left(\lambda+(1-\lambda)\frac{1+\cos\phi_{ij}(t)}{2}\right).
\end{equation}
where $A$ determines the force (in \SI{}{\newton}) and $B$ the range (in \SI{}{\metre}) of repulsive interactions, and $\lambda \in [0,1]$ the (an-)isotropy.
The vector $\mathbf{n}_{ij}(t)$ denotes the normal from $p_i$ to $p_j$ of unit length, $\phi$ the angle between current direction $\mathbf{e}_i(t)$ of particle $p_i$ and $\mathbf{n}_{ij}(t)$.
The scalar value $r_{ij}$ is the sum of the respective radii $r_i$ and $r_j$ of the considered particles, while $d_{ij}(t) = \lVert\mathbf{x}_i(t)-\mathbf{x}_j(t)\rVert$ is the distance between their
centres, both quantities measured in metres. Note that for the experiments performed, the distance $d'_{ij}$ between the perimeters is measured. This is defined as $d'_{ij} = d_{ij}-r_{ij}$.
With regard to the measures taken by the German government for prevention of the spread of COVID-19, we set the critical distance $d_{min}$ for a possible transmission to \SI{1.5}{\metre}.
Figure~\ref{fig:agent_interact} shows different aspects of agent interaction. Figures~\ref{fig:agent_interact_sim}~and~\ref{fig:agent_distance} show interaction in the simulation with regard
to the critical distance, while Figure~\ref{fig:agent_interact_pi_pj} shows associated quantities of the mathematical model between two particles.

The tangential term $\mathbf{f}^{\textrm{tang}}$ (Equation~\ref{equ:force_tang}) is now defined as a fraction $\gamma \in \mathbb{R}$ of the normal term, but in direction $\mathbf{t}_{ij}$, orthogonal to $\mathbf{n}_{ij}$.
Furthermore, this term is added only if the pedestrians involved, $p_i$ and $p_j$, are heading into opposite directions. This is taken care of by the function $\psi_{ij}$ (Equation~\ref{equ:psi}).
\begin{equation}
\label{equ:force_tang}
\mathbf{f}^{\textrm{tang}}(t) = \psi_{ij}(t) \cdot \gamma \cdot \lVert\mathbf{f}^{\textrm{norm}}_{ij}(t)\rVert \cdot \mathbf{t}_{ij}(t).
\end{equation}
The value of $\gamma \in [0,1]$ determines the amount of normal social force added in tangential direction. For the experiments performed, we set $\gamma = 0.7$.
The function $\psi$ (Eq.~\ref{equ:psi}) determines the pedestrians' directions by evaluating the dot product between their respective directions of movement $\mathbf{e}_{i}(t) = \frac{\mathbf{v}_{i}(t)}{\lVert\mathbf{v}_{i}(t)\rVert}$:

\begin{equation}
\label{equ:psi}
\psi_{ij}(t) = \begin{cases}
1 & \langle\mathbf{e}_i(t), \mathbf{e}_j(t)\rangle \leq 0 \\
0 & \, \text{otherwise}\textrm{.}\\
\end{cases}
\end{equation}
The added tangential term makes for a more realistic movement of the agents, as they evade each other early, if applied. This is even more noticeable in conjunction
with the method we employ for pathfinding, as described below.

Physical forces acting between pedestrians are also defined in conformance to Helbing et al:
\begin{equation}
\label{equ:force_ph}
\mathbf{f}^{\textrm{ph}}_{ij}(t) = k \cdot \Theta(r_{ij}-d_{ij}(t)) \cdot \mathbf{n}_{ij}(t) + \kappa \cdot \Theta(r_{ij}-d_{ij}(t)) \cdot \Delta v^{t}_{ji}(t) \cdot \mathbf{t}_{ij}(t).
\end{equation}
Here, constants $\kappa$ and $k$ determine amounts of friction and the force counteracting body compression.
The term $\Delta v^{t}_{ji} = (\mathbf{v}_j - \mathbf{v}_i) \cdot \mathbf{t}_{ij}$ describes the tangential speed difference
between agents $p_i$ and $p_j$. The function $\Theta$ ensures that physical forces only arise when two particles actually touch,
i.\,e.\ the distance $d_{ij}$ between them is lower than the sum of their radii $r_{ij}$:
\begin{equation}
\label{equ:theta}
\Theta(x) = \begin{cases} 0 & \text{if } x < 0 \\ x & \text{otherwise} \textrm{.}\end{cases}
\end{equation}

\begin{figure}[tb]
	\centering
	\begin{subfigure}{0.22\linewidth}
		\includegraphics[height=0.09\textheight]{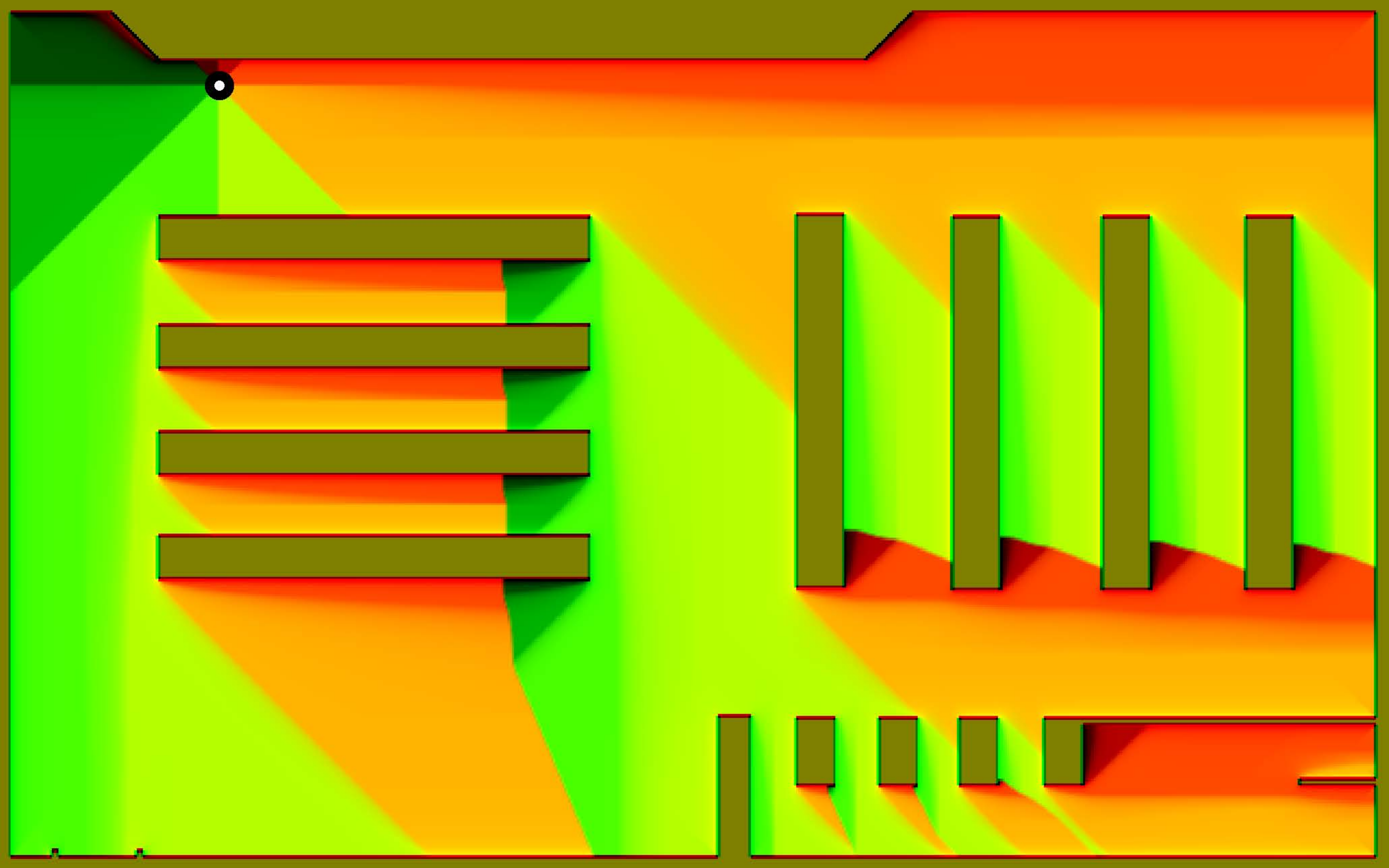}
		\caption{Destination on upper left}
		\label{fig:direction_ul}
	\end{subfigure}
	\hspace{0.15cm}
	\begin{subfigure}{0.22\linewidth}
		\includegraphics[height=0.09\textheight]{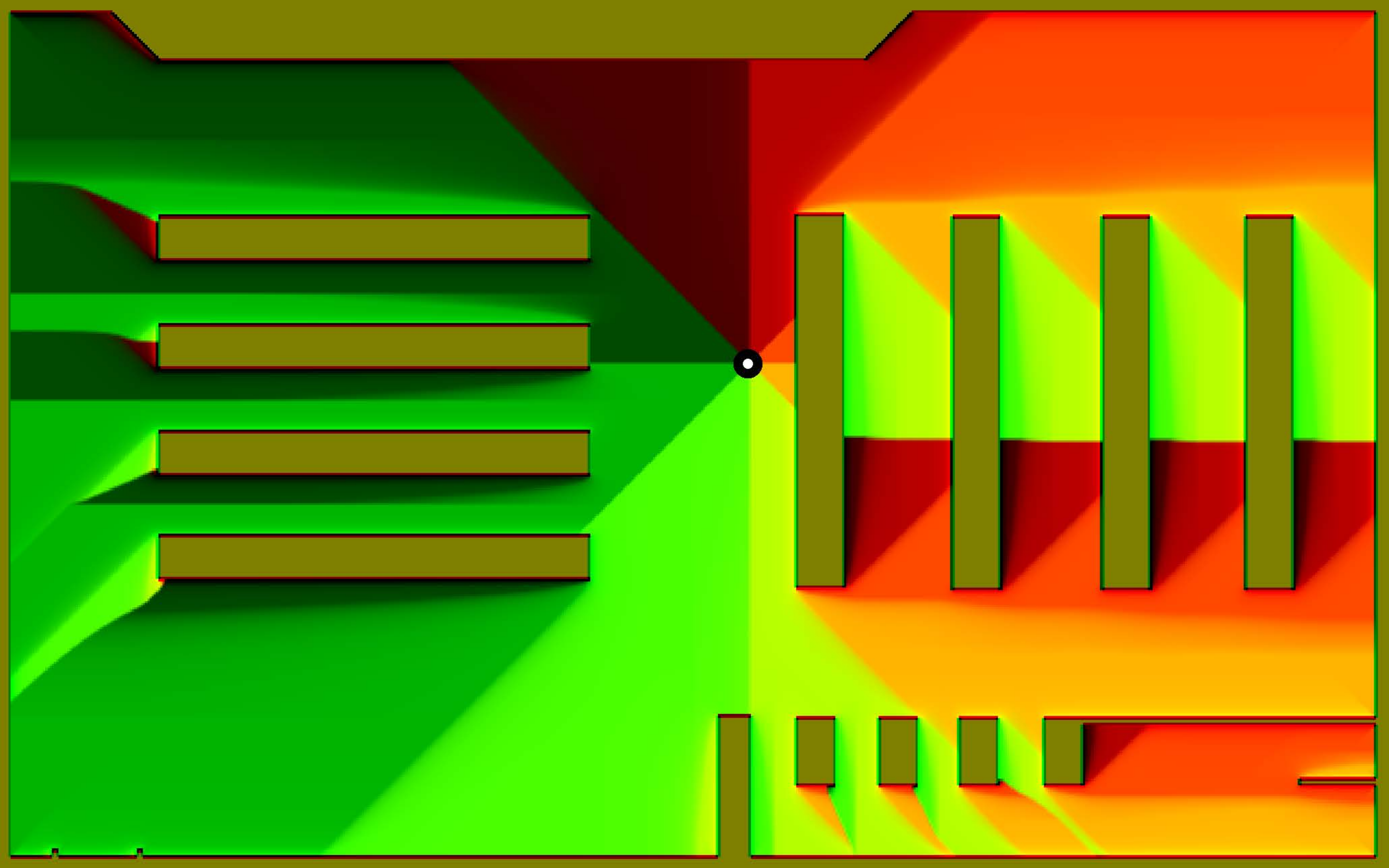}
		\caption{Destination in the center}
		\label{fig:direction_center}
	\end{subfigure}
	\hspace{0.15cm}
	\begin{subfigure}{0.22\linewidth}
		\includegraphics[height=0.09\textheight]{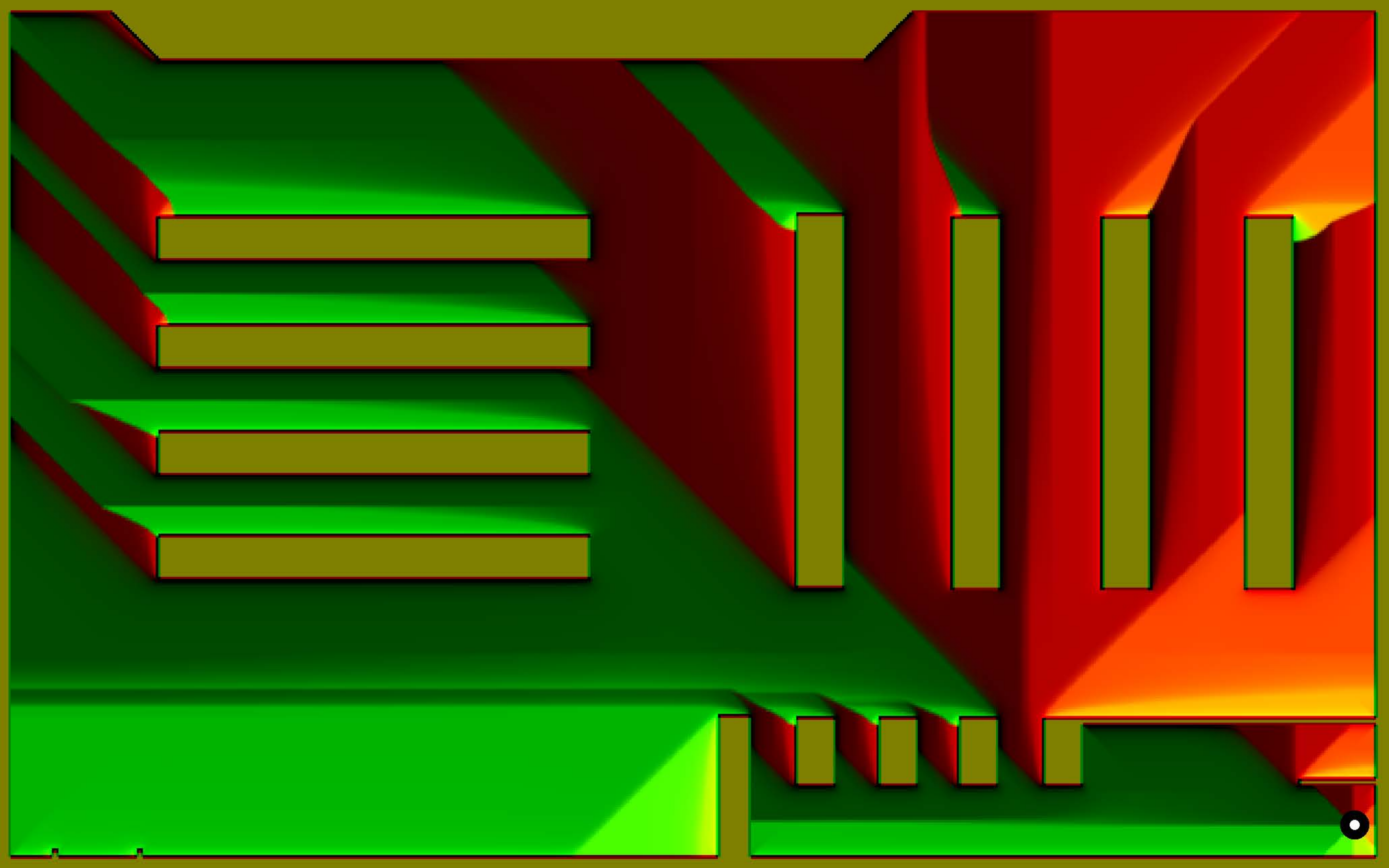}
		\caption{Destination on lower right}
		\label{fig:direction_lr}
	\end{subfigure}
	\hspace{0.15cm}
	\begin{subfigure}{0.22\linewidth}
		\centering
		\includegraphics[height=0.09\textheight]{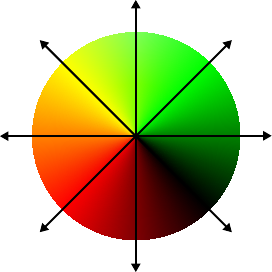}
		\caption{Mapping of colours to directions}
		\label{fig:direction_legend}
	\end{subfigure}
	\caption{Colour-coded direction maps for agent navigation, determining directions of movement for each position within walkable areas (a-c). Destinations are shown as black discs with white
		 dots. Figure (d) shows how colours are mapped to directions}
	\label{fig:direction_map}
\end{figure}

Finally, wall forces $\mathbf{f}^{\textrm{wall}}$ define interaction of agents with obstacles, like walls and stationary objects.
The wall forces are defined in analogy to particle forces:
\begin{equation}
\label{equ:force_wall}
\begin{split}
\mathbf{f}^{\textrm{wall}}_{i}(t) &= \mathbf{f}^{\textrm{wsoc}}_{i}(t) + \mathbf{f}^{\textrm{wph}}_{i}(t) \\
	&= \left( A^{\textrm{wall}} \cdot \exp\left(\frac{r_{i}-d_{ib}(t)}{B^{\textrm{wall}}}\right) + k \cdot \Theta(r_{i}-d_{ib}(t))\right) \cdot \mathbf{n}_{ib}(t) \\
	&- \kappa \cdot \Theta(r_{i}-d_{ib}(t)) \cdot (\mathbf{v}_{i}(t)\cdot\mathbf{t}_{ib}(t)) \cdot \mathbf{t}_{ib}(t).
\end{split}
\end{equation}
In Equation~\ref{equ:force_wall}, $d_{ib}$ denotes the distance, and $\mathbf{n}_{ib}$ the normal and $\mathbf{t}_{ib}$ the tangential direction towards the closest obstacle(-point) $b$ to $p_i$,
which are all determined directly from the scene representation described below.
Parameters $A^{\textrm{wall}}$ and $B^{\textrm{wall}}$ are the analogues to $A$ and $B$ in Equation~\ref{equ:force_norm}.
This is a slight deviation from the model by \emph{Helbing et al.}, as we define separate social distance parameters for obstacles and for other agents.

For the experiments conducted in this work, a scene the agents can move around in needs to be defined. This will define the supermarket scenario considered in the simulation we
performed with regard to the COVID-19 measures, as described in Section~\ref{sec:experiments}.
The scene is represented as a distance transform~\cite{Borgefors:1986, Felzenszwalb:2004} of a two-dimensional, binary discretised map, as depicted in Figure~\ref{fig:supermarket}.
This map defines the size of the simulated area, as well as the regions within which
are walkable (shown as white) or pose an obstacle to the agents (black). The scene in the experiments performed is represented by a discretised map at a resolution of 8 pixels per \SI{}{\metre}.
Agent navigation depends on pre-calculated paths based on \emph{Dijkstra's} algorithm~\cite{Dijkstra:1959}. Accordingly, for each destination, a map containing the directions of movement towards
it for each walkable point in the area is generated. This way the direction of movement $\mathbf{e}_{i}(t)$ of an agent $p_i$ is determined by looking up the given direction of movement at
position $\mathbf{x}_{i}$ from the map corresponding to $p_i$'s current destination $\mathbf{d}_{i}$ (cf.\ Section~\ref{sec:materials_and_methods}).
Colour-coded renderings of maps for three different destination points are shown in Figure~\ref{fig:direction_map}. Agent navigation in the context of our COVID-19 simulation, including details
on how destinations are chosen, is described in Section~\ref{sec:experiments}.

The need for the added tangential social force $\mathbf{f}^{\textrm{tang}}$ becomes apparent in cases of symmetry, which arise when forces are at an equilibrium.
The pre-calculated paths may impose symmetries, as they minimize the distance towards the destination. This means, the prescribed way can be a thin line
even though there is more space available, resulting in unnatural behaviour and building of queues, especially when agents are approaching others head-on.
In the worst case, this can lead to ``deadlocks'' or clogging of pedestrians, even in cases where the surrounding area provides enough
space for the agents to evade each other.

\section{Experiments}
\label{sec:experiments}
\begin{figure}[tb]
	\centering
	\begin{subfigure}[t]{0.45\linewidth}
		\includegraphics[width=\linewidth]{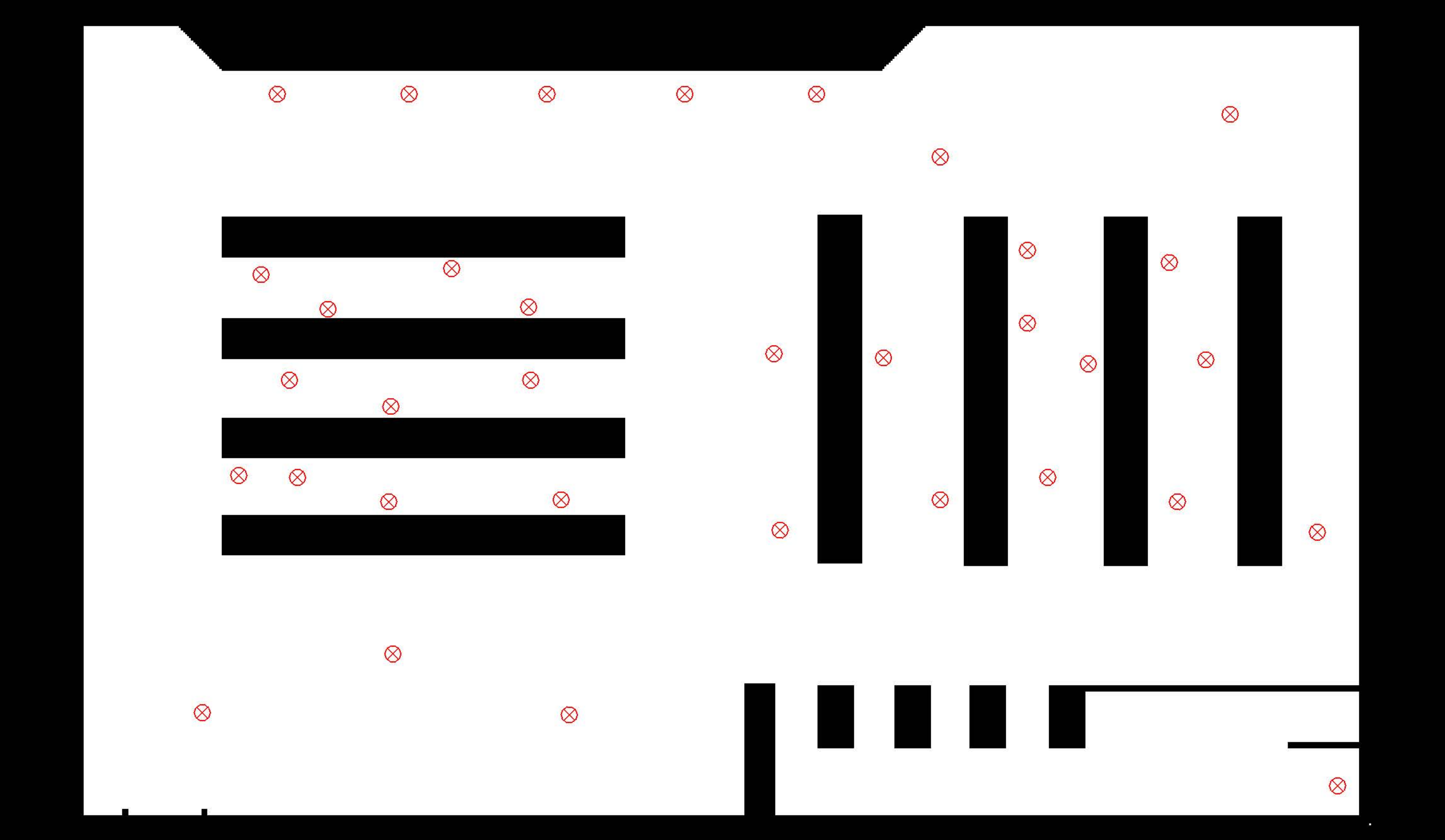}
		\caption{Scene with destinations shown as red circles with crosses inside}
		\label{fig:scene_dest}
	\end{subfigure}
	\hspace{0.5cm}
	\begin{subfigure}[t]{0.45\linewidth}
		\includegraphics[width=\linewidth]{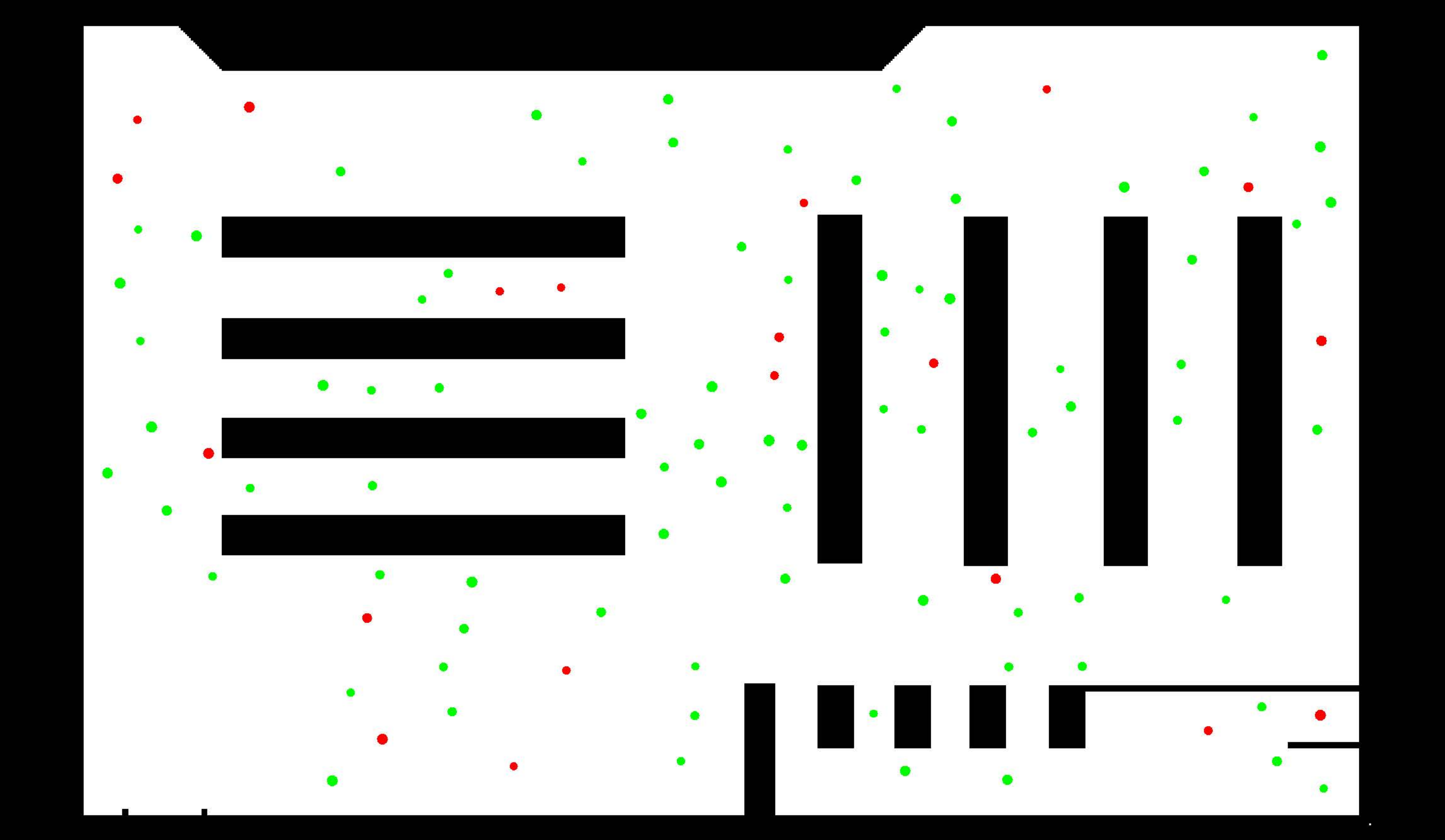}
		\caption{Scene with 100 agents, infected ones shown in red, others in green}
		\label{fig:scene_agents}
	\end{subfigure}
	\caption{Supermarket scene with walkable areas shown in white and obstacles in black}
	\label{fig:supermarket}
\end{figure}

The simulation was carried out with a basic ``supermarket'' scenario of size \SI{80}{\metre} $\times$ \SI{60}{\metre} (cf.\ Figure~\ref{fig:supermarket})
with varying numbers $n$ of agents, which represent customers. The scene is aimed to mimic a typical German supermarket with shelves, counters and cashiers.
Within the simulated area, a set of $34$ destinations is defined (Figure~\ref{fig:scene_dest}), representing points of interest within the supermarket.
We performed simulations for number of agents (population size) of $n \in \{ 50, 100, 200, 300 \}$. A defined amount of the agents were marked as ``infected''.

Figure~\ref{fig:scene_agents} shows
the scene with $100$ agents. Infectious persons are marked red in the visualisation, all others green.
Infected agents carry the virus and can potentially infect others. The amount of infected agents was varied from $\{0.02, 0.05, 0.1, 0.15, 0.2\}$ in the experiments,
tantamount to ratios of $2\%$ to $20\%$.
The agents' radii $r_i$ were sampled uniformly from $[0.25, 0.35]$. Mass is then determined proportionally to the individual radii as $m_i = 160 \cdot r_i$ in \SI{}{\kilogram}.
The desired speed $v_i^0$ was sampled uniformly from the range $[0.3, 0.8]$ (in \SI{}{\metre\second}).

Simulation parameters were chosen as
$A=\SI{10000}{\newton}$,
$B \in \{0.5, 0.7, 1.0, 1.5\} \SI{}{\metre}$,
$A^{\textrm{wall}}= \SI{10000}{\newton}$, $B^{\textrm{wall}} = \SI{0.5}{\metre}$, and $\tau = \SI{0.5}{\second}$.
As body contact does hardly ever occur, constants $k$ and $\kappa$
supposedly have very little impact on the outcome of the
simulation, if at all. For the sake of completeness, we set $k=\SI{20000}{\kilogram\cdot\metre^{-1}\cdot\second^{-1}}$, and $\kappa=\SI{40000}{\kilogram\cdot\second^{-2}}$.

Initially, agents were distributed randomly across the area, i.\,e.\ their starting points $\mathbf{s}_i$ are set to random (walkable) positions within the scene.
The set of destinations were assigned to the agents in an even split, determining the agents' initial destination points $\mathbf{d}_i$.
During the simulation, if an agent reaches its destination, a new destination is assigned randomly from the available set of destinations (excluding the current one).
Agent movement is then determined by the direction map corresponding to the destination, as described in Section~\ref{sec:materials_and_methods}.
Thus, a typical behaviour of customers walking around the supermarket is simulated. The simulation time was set to fifteen minutes.

During the simulation, for each agent, we keep track of exposure time to infected agents. More precisely, the time span an agent comes below the prescribed safety distance $d_{min}$
of $\SI{1.5}{\metre}$ (cf.\ Section \ref{sec:introduction}) to an infected agent is accumulated per individual during the course of the simulation.

\begin{figure}[tb]
	\centering
	\begin{subfigure}[t]{0.45\linewidth}
		\includegraphics[width=\linewidth]{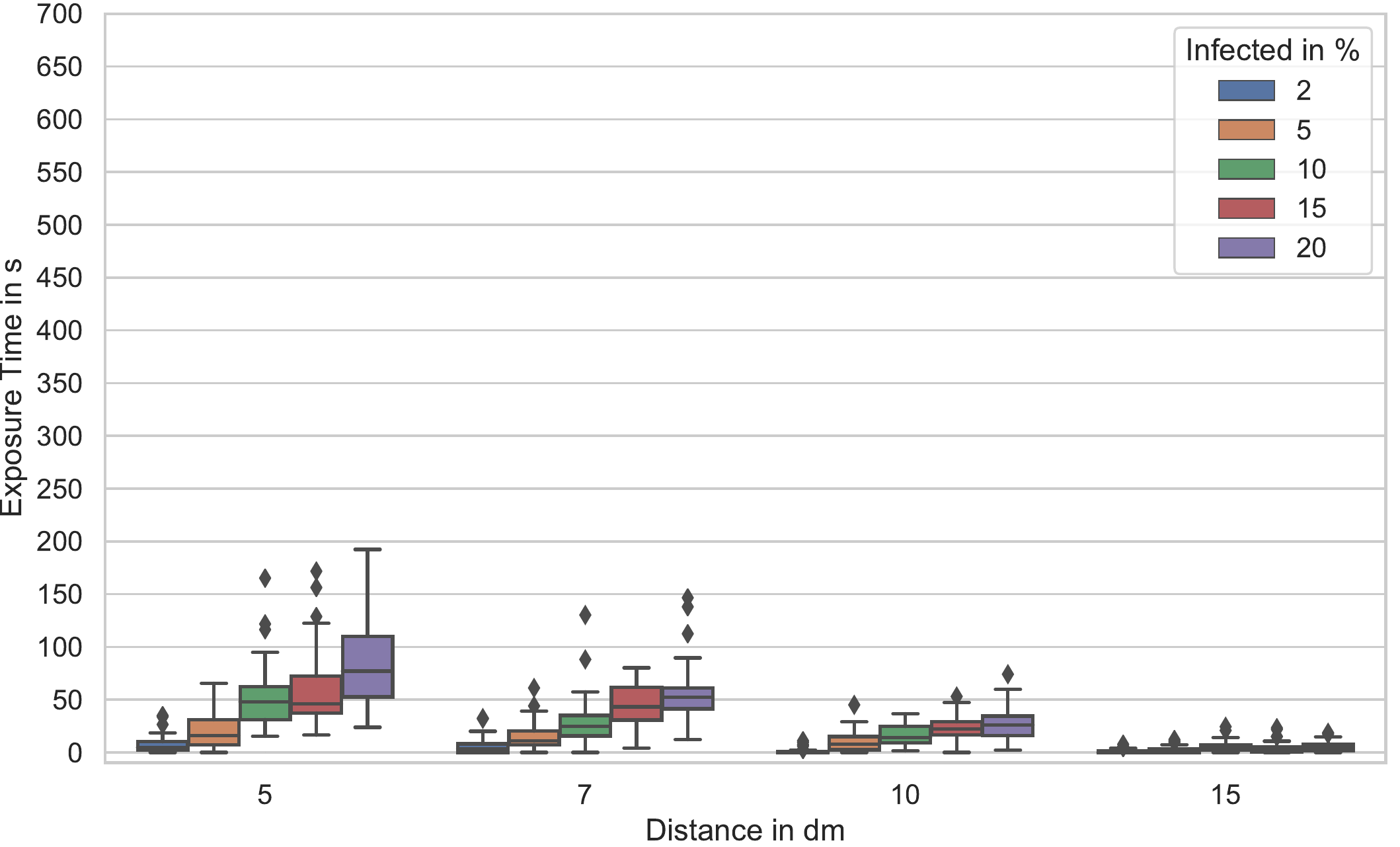}
		\caption{Simulation results for $n = 50$.}
		\label{fig:pop_50}
	\end{subfigure}
	\hspace{0.5cm}
	\begin{subfigure}[t]{0.45\linewidth}
		\includegraphics[width=\linewidth]{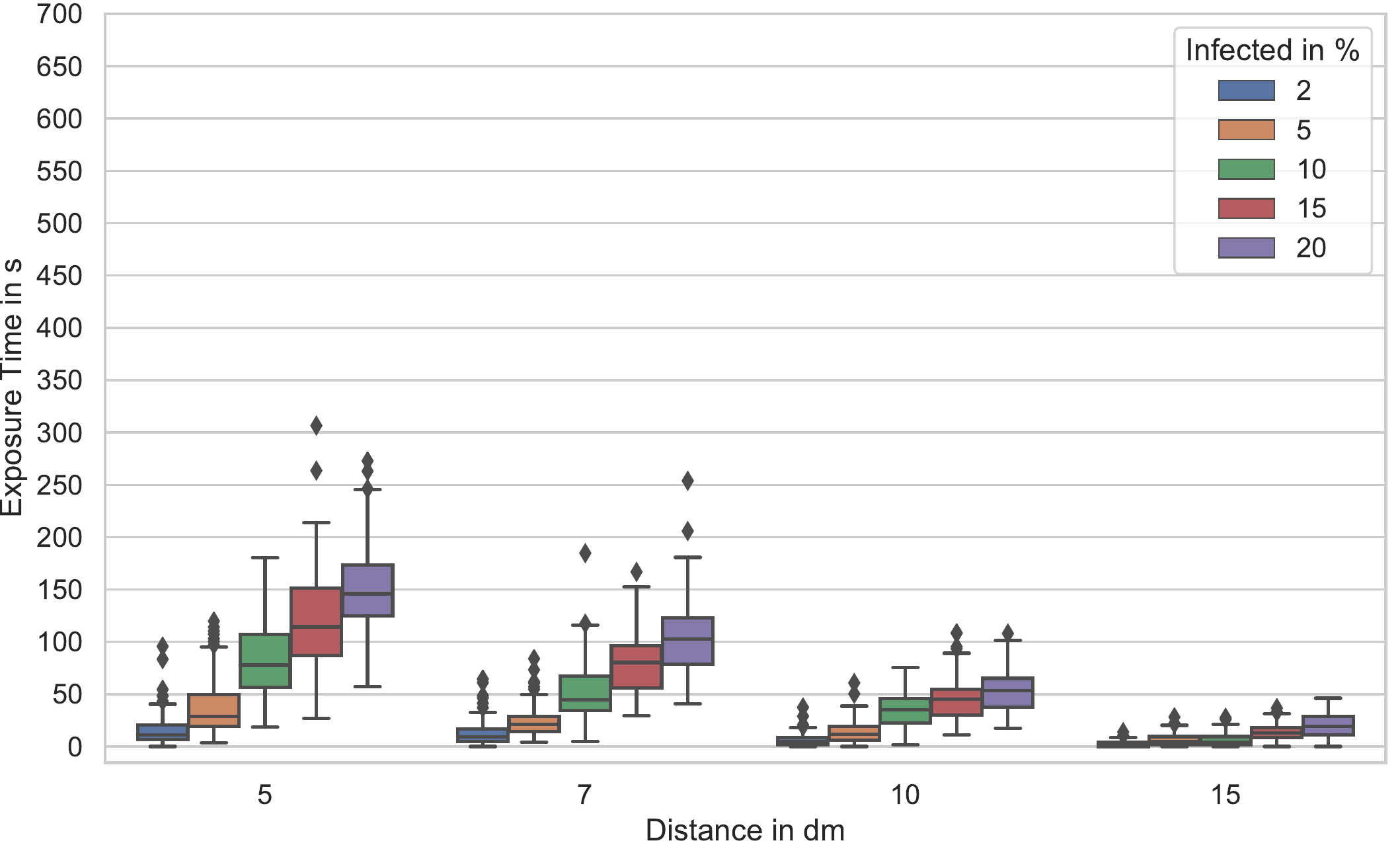}
		\caption{Simulation results for $n = 100$.}
		\label{fig:pop_100}
	\end{subfigure}
	
	\bigskip
	
	\begin{subfigure}[t]{0.45\linewidth}
		\includegraphics[width=\linewidth]{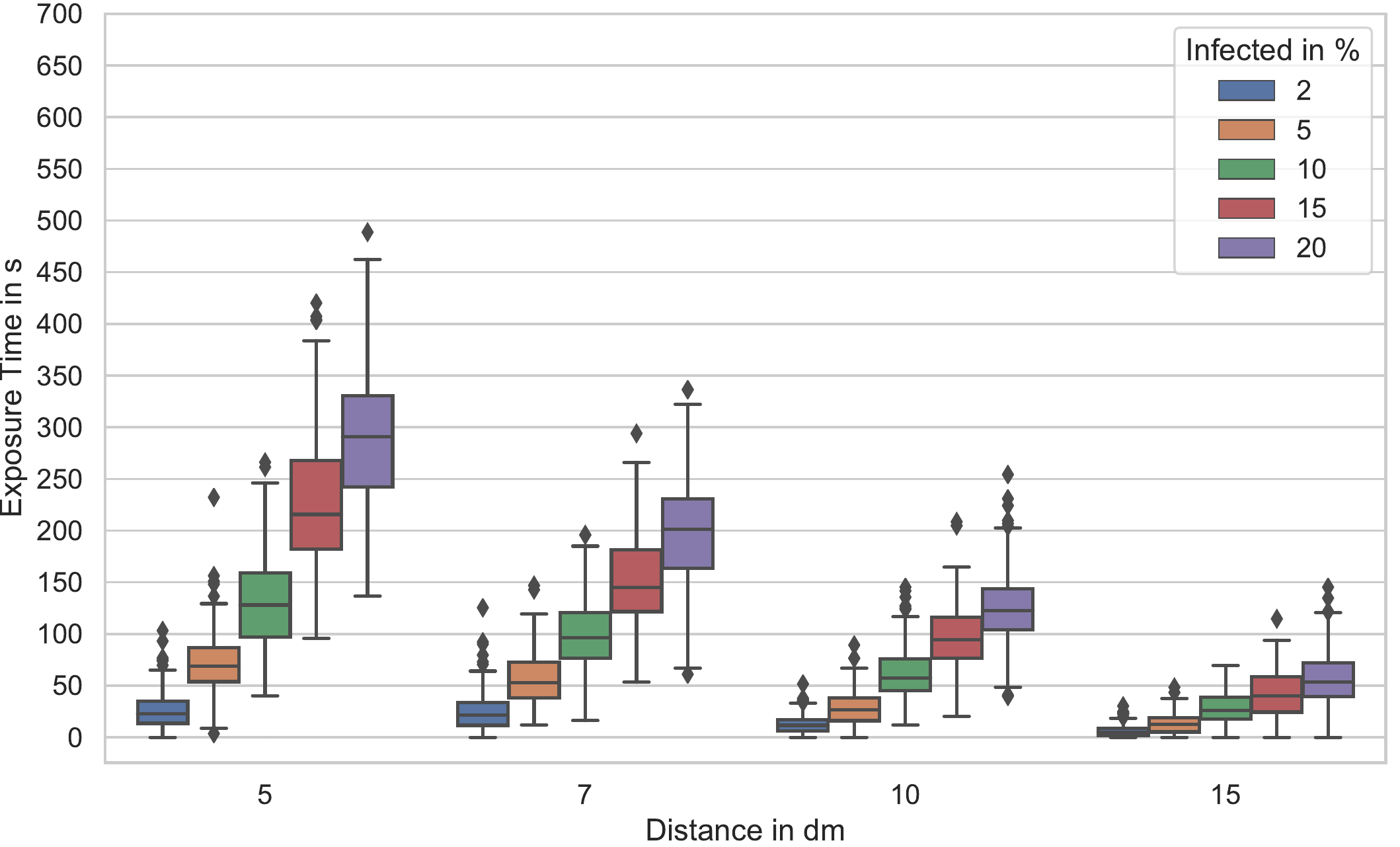}
		\caption{Simulation results $n = 200$.}
		\label{fig:pop_200}
	\end{subfigure}
	\hspace{0.5cm}
	\begin{subfigure}[t]{0.45\linewidth}
		\includegraphics[width=\linewidth]{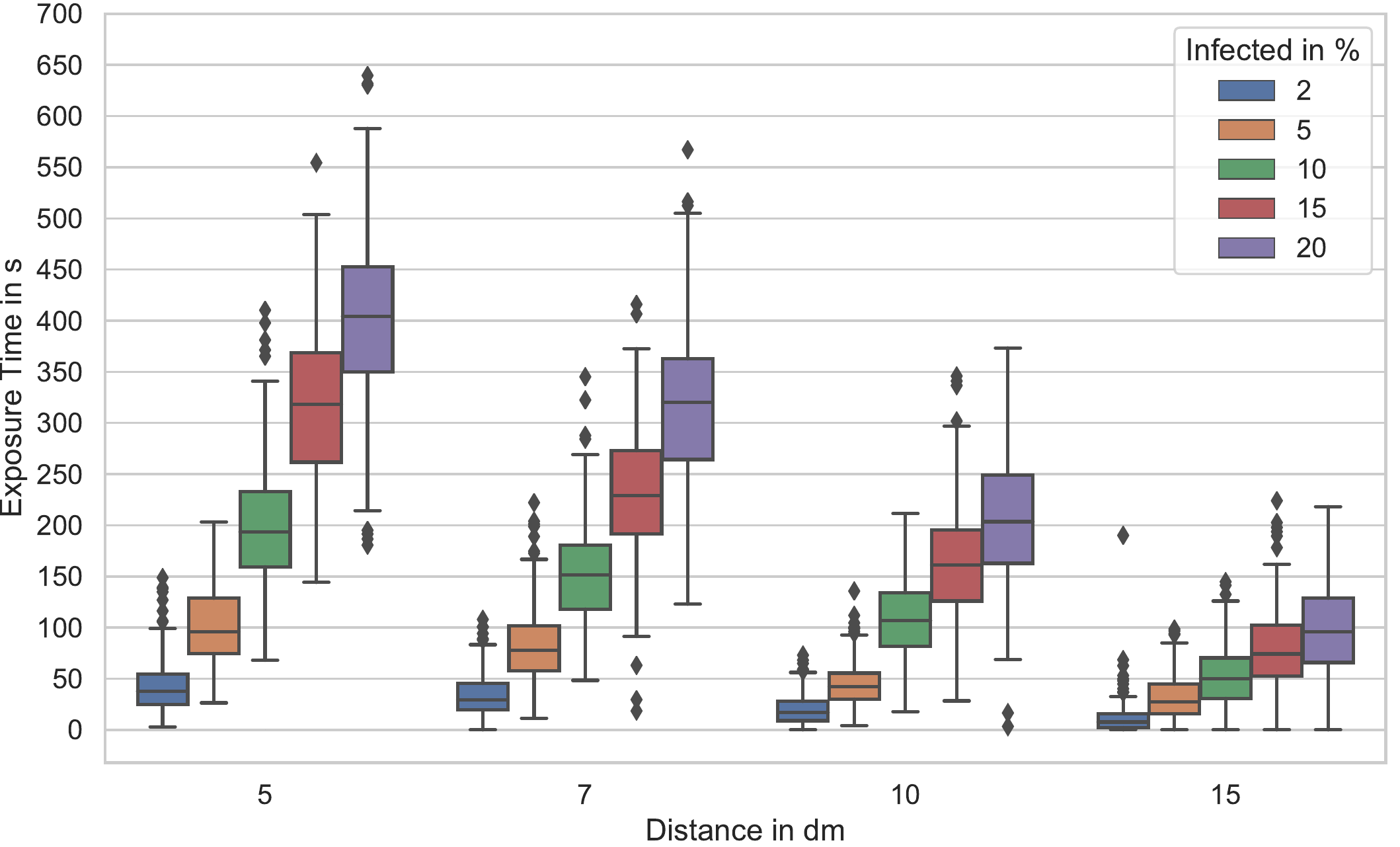}
		\caption{Simulation results $n = 300$.}
		\label{fig:pop_300}
	\end{subfigure}
	\caption{Box-and-whisker plots of the exposure times for populations of $n \in \{50, 100, 200, 300\}$, desired distances of \SI{50}{\centi\metre}, \SI{70}{\centi\metre}, \SI{1}{\metre} and \SI{1.5}{\metre} and infection rates of 2\%, 5\%, 10\%, 15\% and 20\%}
	\label{fig:exposure}
\end{figure}

\section{Results}
\label{sec:results}
The box-and-whisker plots~\cite{Tukey1970} in Figure~\ref{fig:exposure} show the average deviation of the exposure time concerning the different population sizes, distances, and infection rates,
as described in Section~\ref{sec:experiments}. The boxes are representing the interquartile range which contains 50\% of the values and the whiskers are marking the minimum and maximum values, excluding outliers (marked as black diamonds).

The first scenario depicted in Figure~\ref{fig:pop_50} shows the results for a minimal population of $50$ agents. With a uniform distribution of the individuals and if superstructures (cf.\ Figure~\ref{fig:scene_dest}) are neglected a density of one individual per \SI{96}{\square\metre} is to be expected. Thus, enough space for avoidance is available. This is supported by the box plots. Even if a very short desired distance of \SI{50}{\centi\metre} is considered, the mean exposure time is \SI{7.78}{\second} (standard deviation (std) \SI{9.1}{\second}) for an infection rate of 2\% of the individuals. Even if very high infection rates of 10\% are considered the mean exposure time is below one minute (\SI{52}{\second}, std \SI{29.16}{\second}). If the requested distance of \SI{1.5}{\metre} (cf. Section \ref{sec:introduction}) is maintained mean exposure time for an infection rate of 20\% is \SI{2.3}{\second} (std \SI{2.9}{\second}). With a growing number of agents exposure times are rising (cf.\ Figure~\ref{fig:pop_100}). Expected density with a population of $100$ is one individual per \SI{48}{\square\metre}. For small desired distances and medium infection rates of 5\% the mean exposure time is \SI{84.95}{\second} (std \SI{36.5}{\second}) which is 61\% higher than the results with the same parameters and a population of $50$. As the desired distances increase the exposure times are decreasing. Considering the requested \SI{1.5}{\metre} distance the mean exposure time for an infection rate of 20\% is \SI{19.85}{\second} (std \SI{11.79}{\second}), which is 836\% higher than the exposure times for a population of $50$. As far as realistic infection rates of 2\% are concerned, the mean exposure time is \SI{2.49}{\second} (std \SI{2.84}{\second}). The results for a population of size $200$ is shown in Figure~\ref{fig:pop_200}. This population size with an expected density of one agent per \SI{24}{\square\metre} representing the maximum density allowed during lock down in most of German federal states at the time of writing. For an infection rate of 2\% and an desired distance of \SI{1.5}{\metre} the mean exposure time is \SI{6.19}{\second} (std \SI{6.02}{\second}). With an expected density of one individual per \SI{16}{\square\metre} Figure~\ref{fig:pop_300} shows the results for a simulation with $300$ agents. Mean exposure time is \SI{11.31}{\second} (std \SI{15.03}{\second}). The 50\% percentile (median) for all simulations with this parametrisation is below \SI{8}{\second} (\SI{7.7}{\second}, \SI{4.64}{\second}, \SI{1.37}{\second}, \SI{0}{\second}), showing the effectiveness of distancing tactics in the minimisation of exposure times.

\section{Discussion}
\label{sec:discussion}

We have presented a simulation of pedestrian dynamics in realistic scenarios with focus on the spread of infectious diseases by contact transmission.
An important measure taken to reduce the spread of COVID-19 is the so-called \emph{social distancing} or \emph{physical distancing},
aiming to reduce close contacts between individuals in public places.
In the experiments we conducted, we showed how our simulation can give insights about exposure time to infected individuals and the feasibility and effectiveness of keeping
distance in realistic crowded scenarios. Our experiments suggest that, if we assume an infection rate of 2\%, the prescribed minimum distance of \SI{1.5}{\metre} can be maintained
if a density of one person per \SI{16}{\square\metre} is not exceeded.

In this work, we have presented a method for risk assessment concerning pedestrian dynamics and exposure time in conjunction with COVID-19 in particular and infectious diseases in general.
Due to the flexibility of the approach, it can be applied to a great variety of scenarios prone to transmission of contagious diseases. This especially includes public
places, indoor as well as outdoor.

Our simulation can serve as a tool for a better assessment of quantities regarding the number of people to admit, or on guidelines for distances to keep between individual persons.
Due to the nature of the simulation, it can also give insight about optimisation on the geometry of the surrounding, like identification of bottlenecks and hotspots, in
order to reduce risks for people moving around and meeting in the place in question.

With the COVID-19 pandemic affecting countries all over the world at the time of writing, the urgent need of models and tools for better assessment of situations in public places
is apparent. We are confident that our simulation results can serve as a basis for better risk assessment in public places in the context of infectious diseases, and for further research in this area.

\bibliographystyle{unsrt}

\end{document}